# REVISITING STRESS ANALYSIS IN A THREE-DIMENSIONAL ELASTIC HOLLOW SPHERE UNDER UNIAXIAL COMPRESSION VIA THE INVERSE LAPLACE TRANSFORM EXPRESSIONS WITHIN AN ELASTODYNAMIC FRAMEWORK

**S. Takada[a,1] and S. Hokada[b]**

*The stress analysis of a three-dimensional elastic hollow sphere subjected to uniaxial compression is revisited, employing an elastodynamic framework. Through the application of the Laplace transform, the scalar and vector potentials of displacement are expanded, facilitating a detailed exploration of the system's mechanical behavior. The static solutions for displacement and stress distributions are derived in the long-time limit, which reveal key insights into the response of the elastic hollow sphere. Notably, on the inner surface, certain quantities exhibit a peak at the point where the angle between the compressive force and the point on the surface becomes perpendicular, indicating localized stress concentration. These findings provide a robust analytical approach for understanding and predicting the behavior of elastic hollow spheres under uniaxial loading, with implications for material science and structural engineering.*

**Introduction.** The stress analysis of elastic spheres under compression has long been a cornerstone of elasticity theory [1–4], with applications spanning material science, engineering design, and structural analysis. A solid sphere subjected to uniaxial compression provides a canonical test case for understanding stress and strain distributions in two- and three-dimensional elastic bodies. Theoretical studies have meticulously clarified the behavior of stress along the load axis, offering insights into deformation patterns, stress concentrations, and potential failure mechanisms [1–3]. These findings have been extensively documented in classical textbooks [1–4] and research literature [5–7], forming the basis for numerous practical applications.

In addition to static loading scenarios, elastodynamic theory has been applied to analyze the behavior of cylinders or spheres under dynamic or impact loads [8–12]. Such investigations are particularly relevant in fields like ballistics, geomechanics, and composite material testing, where time-dependent loading plays a critical role. These studies have revealed complex stress wave propagation phenomena and dynamic stress concentration effects, further enriching our understanding of spherical elastic bodies.

However, real-world materials and components are rarely idealized as perfectly solid. Most materials inherently exhibit imperfections such as voids, pores, or inclusions, which can significantly alter their mechanical response [5, 13, 14]. For instance, internal porosity is a common feature in ceramics, sintered metals, and biological materials such as bone. These pores affect the stress distribution within the material, often acting as stress concentrators that influence deformation, fracture behavior, and overall structural integrity.

While earlier studies [5, 13] have addressed the mechanics of hollow and porous cylinders or spheres, there remains a notable gap in the literature concerning comprehensive and explicit solutions for such problems. A seminal study by Kasano *et al.* [13] tackled the stress analysis of a hollow sphere under uniaxial compression, considering the effects of internal voids. Their approach, however, determined the coefficients of displacement and stress components in an incremental form, which, while insightful, lacked the clarity and utility of explicit closed-form expressions.

[a]Department of Mechanical Systems Engineering, Tokyo University of Agriculture and Technology, Tokyo, Japan ([1]takada@go.tuat.ac.jp). [b]Department of Industrial Technology and Innovation, Tokyo University of Agriculture and Technology, Tokyo, Japan.

This study revisits the problem of stress analysis in a three-dimensional elastic hollow sphere under uniaxial compression to address this gap. By leveraging advancements in analytical techniques and computational tools, we aim to derive explicit expressions for the displacement and stress coefficients. These results not only enhance the theoretical understanding of stress behavior in hollow spheres but also provide practical tools for predicting the mechanical performance of porous materials.

The explicit determination of these coefficients has far-reaching implications. It simplifies the computational modeling of porous structures, aids in the design of materials with tailored mechanical properties, and improves the accuracy of failure predictions in engineering applications. Furthermore, this work contributes to broader efforts in understanding and optimizing the behavior of elastic bodies with complex internal geometries, paving the way for innovations in materials science, biomechanics, and structural engineering.

The organization of this paper is as follows: In the next Section, we explain the model and the setup. In Sect. 2, the elastodynamic theory is applied to derive the expressions of the displacement and the stress. In Sect. 3, the explicit expressions of the static solutions are derived in the long-time limit. In Sect. 4, we present various results obtained from the static solutions. In the last section, we conclude and discuss our results.

**1. Model and Setup.** Let us consider an elastic hollow sphere whose outer and inner radii are given by $R_\mathrm{o}$ and $R_\mathrm{i}$, respectively. We also assume that the mass density, the shear modulus and Poisson's ratio of the sphere are given by $\varrho_0$, $G$ and $\nu$, respectively. Now, we add the stress acting on the sphere diametrically as shown in Fig. 1.

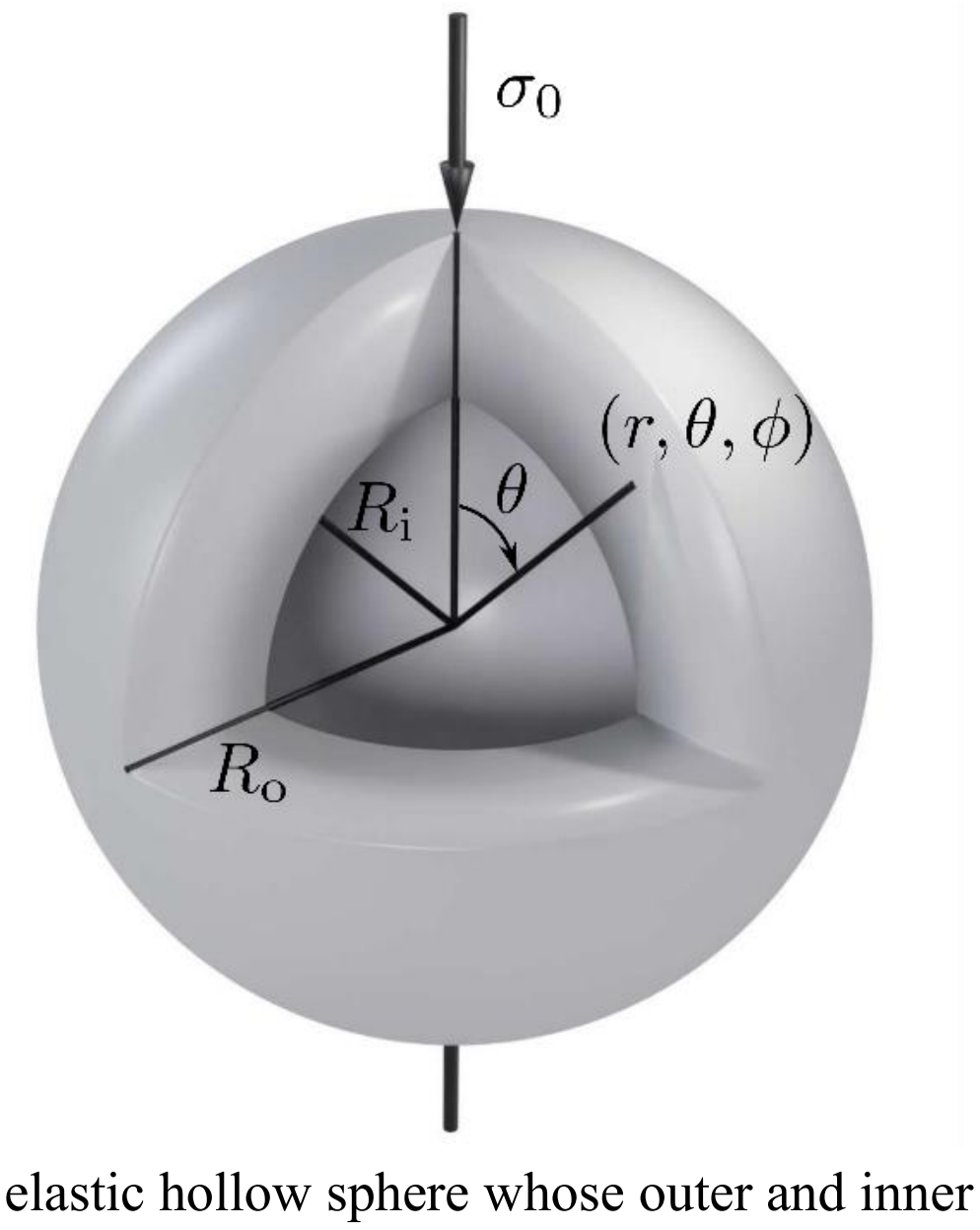

Fig. 1. A schematic of the system. An elastic hollow sphere whose outer and inner radii are $R_\mathrm{o}$ and $R_\mathrm{i}$, respectively, is compressed uniaxially ($\theta = 0$ and $\pi$) with the pressure $\sigma_0$.

The boundary conditions are given by

$$\sigma_{rr}|_{r=R_\mathrm{o}} = -\sigma_0 \Theta(t) \sum_{n=0,2,4,\cdots}^{\infty} (2n+1) P_n(\cos\theta)\,, \tag{1a}$$

$$\sigma_{rr}|_{r=R_\mathrm{i}} = \sigma_{r\theta}|_{r=R_\mathrm{o}} = \sigma_{r\theta}|_{r=R_\mathrm{i}} = 0, \tag{1b}$$

where $P_n(x)$ stands for the Legendre polynomial [15]. It is convenient to introduce the following scaled (dimensionless) parameters:

$$\boldsymbol{\rho} \equiv \frac{\boldsymbol{r}}{R_0}, \qquad \tilde{\boldsymbol{u}} \equiv \frac{G}{\sigma_0}\frac{\boldsymbol{u}}{R_0}, \qquad \tau \equiv \frac{v_{\mathrm{L}} t}{R_0}, \qquad \overleftrightarrow{\tilde{\sigma}} \equiv \frac{\tilde{\sigma}}{\sigma_0}, \tag{2}$$

where $v_{\mathrm{L}} \equiv \sqrt{2(1-\nu)G/[(1-2\nu)\varrho_0]}$ is the longitudinal wave velocity of the sphere.

**2. Elastodynamic theory.** Under the situation explained in the previous section, we solve the following dimensionless Navier-Cauchy equation [2]:

$$\mu^2 \frac{\partial^2}{\partial \tau^2} \tilde{\boldsymbol{u}} = \nabla_\rho^2 \tilde{\boldsymbol{u}} + \frac{1}{1-2\nu} \boldsymbol{\nabla}_\rho (\boldsymbol{\nabla}_\rho \cdot \tilde{\boldsymbol{u}}), \tag{3}$$

with $\boldsymbol{\nabla}_\rho \equiv R_0 \boldsymbol{\nabla}$ and the speed ratio of the primary to the secondary wave given by

$$\mu \equiv \sqrt{\frac{2(1-\nu)}{1-2\nu}}. \tag{4}$$

Let us consider the Helmholtz theorem:

$$\tilde{\boldsymbol{u}} = \boldsymbol{\nabla}_\rho \varphi(\rho, \theta) + \boldsymbol{\nabla}_\rho \times \boldsymbol{\nabla}_\rho \times \rho \chi(\rho, \theta) \boldsymbol{e}_\rho, \tag{5}$$

where $\varphi$ and $\chi$ correspond to scalar and vector potentials, respectively, and $\boldsymbol{e}_\rho$ is the unit vector in the $\rho$-direction. By substituting Eq. (5) into the Navier-Cauchy equation (3), we find that the two potentials should satisfy the wave equations with the speeds 1 and $1/\mu$, respectively. Now, it is convenient to introduce the Laplace transforms of the potentials as

$$\begin{Bmatrix} \mathcal{L}[\varphi](\rho, \theta, s) \\ \mathcal{L}[\chi](\rho, \theta, s) \end{Bmatrix} \equiv \int_0^\infty \begin{Bmatrix} \varphi(\rho, \theta, \tau) \\ \chi(\rho, \theta, \tau) \end{Bmatrix} \mathrm{e}^{-s\tau} \mathrm{d}\tau. \tag{6}$$

Then, we find that these two quantities are expanded as

$$\mathcal{L}[\varphi](\rho, \theta, s) = \sum_{n=0,2,4,\cdots}^{\infty} [a_n i_n(\rho s) + b_n k_n(\rho s)] P_n(\cos\theta), \tag{7a}$$

$$\mathcal{L}[\chi](\rho, \theta, s) = \sum_{n=0,2,4,\cdots}^{\infty} [c_n i_n(\mu \rho s) + d_n k_n(\mu \rho s)] P_n(\cos\theta), \tag{7b}$$

where the coefficients $a_n$, $b_n$, $c_n$, and $d_n$ should be determined by the boundary condition at the surface given by Eq. (1). Here, $i_n(s)$ and $k_n(s)$ are the modified spherical Bessel functions of the first and second kinds, respectively, and given by

$$i_n(s) \equiv \sqrt{\frac{\pi}{2s}} I_{n+1/2}(s), \qquad k_n(s) \equiv \sqrt{\frac{2}{\pi s}} K_{n+1/2}(s), \tag{8}$$

with the modified Bessel functions of the first and second kinds [15]

$$I_n(s) \equiv \sum_{m=0}^{\infty} \frac{1}{m!\,\Gamma(m+n+1)} \left(\frac{s}{2}\right)^{2m+n}, \tag{9a}$$

$$K_n(s) \equiv \frac{\pi}{2} \frac{I_{-n}(s) - I_n(s)}{\sin(n\pi)}, \tag{9b}$$

as well as the Gamma function $\Gamma(s) = \int_0^\infty t^{s-1} \mathrm{e}^{-t} \mathrm{d}t$ [15].

Using these two potentials, each component of the displacement is evaluated by Eq. (5). Then, each component of the stress can be obtained in terms of the combination of the (dimensionless) stress-strain relation for linear elasticity

$$\tilde{\sigma}_{\alpha\beta} = \frac{2\nu}{1-2\nu} \delta_{\alpha\beta} \tilde{\varepsilon}_{\gamma\gamma} + 2\tilde{\varepsilon}_{\alpha\beta}, \tag{10}$$

and the strain-displacement relations in the polar coordinates, where $\tilde{\varepsilon}_{\alpha\beta} \equiv (G/\sigma_0)\varepsilon_{\alpha\beta}$ is the $\alpha\beta$-component of the scaled strain.

The coefficients that appear in Eqs. (7) need to satisfy the boundary conditions (1), where the orthogonality of the Legendre polynomials is used. This yields simultaneous equations, which determine the expressions of the coefficients. The explicit expressions of the coefficients are given by

$$\begin{Bmatrix} a_n \\ b_n \\ c_n \\ d_n \end{Bmatrix} = \begin{Bmatrix} - \\ + \\ - \\ + \end{Bmatrix} \frac{2n+1}{2} \frac{1}{sD_n(s)} \begin{Bmatrix} N_{n,1}(s) \\ N_{n,2}(s) \\ N_{n,3}(s) \\ N_{n,4}(s) \end{Bmatrix}, \tag{11}$$

where the coefficients $D_n(s)$ and $N_{n,i}(s)$ $(i = 1, \cdots, 4)$ are listed in Table 1.

Table 1. List of the expressions of $D_n(s)$ and $N_{n,i}(s)$ $(i = 1, \cdots, 4)$ with the associated functions $F_{n,i}(s)$ and $G_{n,i}(s)$ $(i = 1, 2, 3)$.

| |
|---|
| $F_{n,1}(s) = \left[n(n-1) + \frac{1-\nu}{1-2\nu}s^2\right] i_n(s) - 2s i_{n+1}(s)$ <br> $F_{n,2}(s) = (n-1)i_n(s) + s i_{n+1}(s)$ <br> $F_{n,3}(s) = \left(n^2 - 1 + \frac{s^2}{2}\right) i_n(s) - s i_{n+1}(s)$ |
| $G_{n,1}(s) = \left[n(n-1) + \frac{1-\nu}{1-2\nu}s^2\right] k_n(s) + 2s k_{n+1}(s)$ <br> $G_{n,2}(s) = (n-1)k_n(s) - s k_{n+1}(s)$ <br> $G_{n,3}(s) = \left(n^2 - 1 + \frac{s^2}{2}\right) k_n(s) + s k_{n+1}(s)$ |
| $D_n(s) = \begin{vmatrix} F_{n,1}(s) & G_{n,1}(s) & n(n+1)F_{n,2}(\mu s) & n(n+1)G_{n,2}(\mu s) \\ F_{n,1}(\rho_\mathrm{i} s) & G_{n,1}(\rho_\mathrm{i} s) & (n+1)F_{n,2}(\mu\rho_\mathrm{i} s) & n(n+1)G_{n,2}(\mu\rho_\mathrm{i} s) \\ F_{n,2}(s) & G_{n,2}(s) & F_{n,3}(\mu s) & G_{n,3}(\mu s) \\ F_{n,2}(\rho_\mathrm{i} s) & G_{n,2}(\rho_\mathrm{i} s) & F_{n,3}(\mu\rho_\mathrm{i} s) & G_{n,3}(\mu\rho_\mathrm{i} s) \end{vmatrix}$ |
| $N_{n,1}(s) = \begin{vmatrix} G_{n,1}(\rho_\mathrm{i} s) & (n+1)F_{n,2}(\mu\rho_\mathrm{i} s) & (n+1)G_{n,2}(\mu\rho_\mathrm{i} s) \\ G_{n,2}(s) & F_{n,3}(\mu s) & G_{n,3}(\mu s) \\ G_{n,2}(\rho_\mathrm{i} s) & F_{n,3}(\mu\rho_\mathrm{i} s) & G_{n,3}(\mu\rho_\mathrm{i} s) \end{vmatrix}$ |
| $N_{n,2}(s) = \begin{vmatrix} F_{n,1}(\rho_\mathrm{i} s) & (n+1)F_{n,2}(\mu\rho_\mathrm{i} s) & n(n+1)G_{n,2}(\mu\rho_\mathrm{i} s) \\ F_{n,2}(s) & F_{n,3}(\mu s) & G_{n,3}(\mu s) \\ F_{n,2}(\rho_\mathrm{i} s) & F_{n,3}(\mu\rho_\mathrm{i} s) & G_{n,3}(\mu\rho_\mathrm{i} s) \end{vmatrix}$ |
| $N_{n,3}(s) = \begin{vmatrix} F_{n,1}(\rho_\mathrm{i} s) & G_{n,1}(\rho_\mathrm{i} s) & n(n+1)G_{n,2}(\mu\rho_\mathrm{i} s) \\ F_{n,2}(s) & G_{n,2}(s) & G_{n,3}(\mu s) \\ F_{n,2}(\rho_\mathrm{i} s) & G_{n,2}(\rho_\mathrm{i} s) & G_{n,3}(\mu\rho_\mathrm{i} s) \end{vmatrix}$ |
| $N_{n,4}(s) = \begin{vmatrix} F_{n,1}(\rho_\mathrm{i} s) & G_{n,1}(\rho_\mathrm{i} s) & (n+1)F_{n,2}(\mu\rho_\mathrm{i} s) \\ F_{n,2}(s) & G_{n,2}(s) & F_{n,3}(\mu s) \\ F_{n,2}(\rho_\mathrm{i} s) & G_{n,2}(\rho_\mathrm{i} s) & F_{n,3}(\mu\rho_\mathrm{i} s) \end{vmatrix}$ |

By calculating the inverse Laplace transform, $\tilde{u}_\alpha$ and $\tilde{\sigma}_{\alpha\beta}$ are given by

$$\tilde{u}_\rho = -\sum_{n=0,2,4,\cdots}^{\infty} \frac{2n+1}{2} \mathcal{L}^{-1}\left[\frac{N_{n,1}(s)}{\rho s D_n(s)}\right] P_n(\cos\theta), \tag{12a}$$

$$\tilde{u}_\theta = \sum_{n=0,2,4,\cdots}^{\infty} \frac{2n+1}{2} \mathcal{L}^{-1}\left[\frac{N_{n,2}(s)}{\rho s D_n(s)}\right] \sin\theta\, P_n'(\cos\theta), \tag{12b}$$

$$\tilde{\sigma}_{\rho\rho} = -\sum_{n=0,2,4,\cdots}^{\infty} (2n+1) \mathcal{L}^{-1}\left[\frac{N_{n,3}(s)}{\rho^2 s D_n(s)}\right] P_n(\cos\theta), \tag{12c}$$

$$\tilde{\sigma}_{\rho\theta} = \sum_{n=0,2,4,\cdots}^{\infty} (2n+1)\mathcal{L}^{-1}\left[\frac{N_{n,4}(s)}{\rho^2 s D_n(s)}\right]\sin\theta\, P_n'(\cos\theta)\,, \tag{12d}$$

$$\tilde{\sigma}_{\theta\theta} = \sum_{n=0,2,4,\cdots}^{\infty} (2n+1)\left\{\mathcal{L}^{-1}\left[\frac{N_{n,5}(s)}{\rho^2 s D_n(s)}\right]P_n(\cos\theta) - \mathcal{L}^{-1}\left[\frac{N_{n,2}(s)}{\rho^2 s D_n(s)}\right]\cos\theta\, P_n'(\cos\theta)\right\}, \tag{12e}$$

$$\tilde{\sigma}_{\phi\phi} = -\sum_{n=0,2,4,\cdots}^{\infty} (2n+1)\left\{\mathcal{L}^{-1}\left[\frac{N_{n,6}(s)}{\rho^2 s D_n(s)}\right]P_n(\cos\theta) - \mathcal{L}^{-1}\left[\frac{N_{n,2}(s)}{\rho^2 s D_n(s)}\right]\cos\theta\, P_n'(\cos\theta)\right\}, \tag{12f}$$

where $\mathcal{L}^{-1}[f] = 1(2\pi/\mathrm{i})\int_{\gamma-\mathrm{i}\infty}^{\gamma+\mathrm{i}\infty} f(s)\mathrm{e}^{st}\mathrm{d}s$ indicates the inverse Laplace transform of $f$, where $\gamma$ (>0) should be larger than any real parts of the poles, and $P_n'(\cos\theta) = \mathrm{d}P_n(x)/\mathrm{d}x|_{x=\cos\theta}$. In general, the computation of the Laplace transforms is time-consuming. Previous studies treated this problem by using the technique of the fast Fourier transform [16] or reducing to the computation of the residues of the integrands [8, 9, 11, 12].

**3. Static solutions.** In this paper, we focus on explicit expressions of the static condition. These expressions are obtained in the long-time limits of Eqs. (12) and corresponding expressions of the other quantities. For this purpose, we can use the final value theorem: $f^{(\mathrm{st})} \equiv \lim_{\tau\to\infty} f(\tau) = \lim_{s\to 0}\{s\mathcal{L}[f](s)\}$. After some calculations, one gets

$$\tilde{u}_\rho^{(\mathrm{st})}(\rho,\theta) = -\sum_{n=0,2,4,\cdots}^{\infty} \frac{2n+1}{2(n-1)(n+2)}\frac{\mathcal{N}_n^{(1)}}{\mathcal{D}_n}\rho^{n-1}P_n(\cos\theta)\,, \tag{13a}$$

$$\tilde{u}_\theta^{(\mathrm{st})}(\rho,\theta) = \sum_{n=0,2,4,\cdots}^{\infty} \frac{2n+1}{2(n-1)(n+2)}\frac{\mathcal{N}_n^{(2)}}{\mathcal{D}_n}\rho^{n-1}\sin\theta\, P_n'(\cos\theta)\,, \tag{13b}$$

$$\tilde{\sigma}_{\rho\rho}^{(\mathrm{st})}(\rho,\theta) = -\sum_{n=0,2,4,\cdots}^{\infty} \frac{2n+1}{(n-1)(n+2)}\frac{\mathcal{N}_n^{(3)}}{\mathcal{D}_n}\rho^{n-2}P_n(\cos\theta)\,, \tag{13c}$$

$$\tilde{\sigma}_{\rho\theta}^{(\mathrm{st})}(\rho,\theta) = \sum_{n=0,2,4,\cdots}^{\infty} \frac{2n+1}{(n-1)(n+2)}\frac{\mathcal{N}_n^{(4)}}{\mathcal{D}_n}\rho^{n-2}\sin\theta\, P_n'(\cos\theta)\,, \tag{13d}$$

$$\tilde{\sigma}_{\theta\theta}^{(\mathrm{st})}(\rho,\theta) = \sum_{n=0,2,4,\cdots}^{\infty} \frac{2n+1}{(n-1)(n+2)}\rho^{n-2}\left[\frac{\mathcal{N}_n^{(5)}}{\mathcal{D}_n}P_n(\cos\theta) - \frac{\mathcal{N}_n^{(2)}}{\mathcal{D}_n}\cos\theta\, P_n'(\cos\theta)\right], \tag{13e}$$

$$\tilde{\sigma}_{\phi\phi}^{(\mathrm{st})}(\rho,\theta) = -\sum_{n=0,2,4,\cdots}^{\infty} \frac{2n+1}{(n-1)(n+2)}\rho^{n-2}\left[\frac{\mathcal{N}_n^{(6)}}{\mathcal{D}_n}P_n(\cos\theta) - \frac{\mathcal{N}_n^{(2)}}{\mathcal{D}_n}\cos\theta\, P_n'(\cos\theta)\right], \tag{13f}$$

where the expressions of $\mathcal{D}_n$ and $\mathcal{N}_n^{(i)}$ $(i = 1,\cdots,6)$ are listed in Table 2.

Table 2. List of the expressions of $\mathcal{D}_n$ and $\mathcal{N}_n^{(i)}$ $(i = 1,\cdots,6)$ with the associated coefficients $\mathcal{C}_n^{(i)}$ $(i = 1,\cdots,4)$.

| |
|---|
| $\mathcal{C}_n^{(1)} = (n+2)\{(2n+1)[(n-1)n(n+1)(n+2) + 4(1-\nu^2)](1-\rho_\mathrm{i}^{2n+3}) - (n-1)(n+1)(2n+3)(n^2-2+2\nu)(1-\rho_\mathrm{i}^{2n+1})\}$ |
| $\mathcal{C}_n^{(2)} = (n-1)(n+2)[n(n+2)(2n-1)(1-\rho_\mathrm{i}^{2n+1}) - (2n+1)(n^2-2+2\nu)(1-\rho_\mathrm{i}^{2n-1})]$ |
| $\mathcal{C}_n^{(3)} = (n-1)\{2[(n^2-2)(n^2+n+1) + n(n+2)(2n-1)\nu](1-\rho_\mathrm{i}^{2n+1}) + (n-1)n(n+1)(n+2)(2n+1)(1-\rho_\mathrm{i}^2) + 4(2n+1)[(1-\rho_\mathrm{i}^2) + \nu^2\rho_\mathrm{i}^2(1-\rho_\mathrm{i}^{2n-1})]\}$ |
| $\mathcal{C}_n^{(4)} = (n-1)(n+2)\{2[n^2+n+1+(2n+1)\nu](1-\rho_\mathrm{i}^{2n+3}) + (n-1)(n+1)(2n+3)(1-\rho_\mathrm{i}^2)\}$ |
| $\mathcal{D}_n = 4[(n^2+n+1)^2 - (2n+1)^2\nu^2](1-\rho_\mathrm{i}^{2n-1})(1-\rho_\mathrm{i}^{2n+3}) - (n-1)n(n+1)(n+2)(2n-1)(2n+3)\rho_\mathrm{i}^{2n-1}(1-\rho_\mathrm{i}^2)^2$ |
| $\mathcal{N}_n^{(1)} = n\mathcal{C}_n^{(1)} - (n+1)(n-2+4\nu)\rho^2\mathcal{C}_n^{(2)} - (n+1)\left(\frac{\rho_\mathrm{i}}{\rho}\right)^{2n+1}\mathcal{C}_n^{(3)} + n(n+3-4\nu)\left(\frac{\rho_\mathrm{i}}{\rho}\right)^{2n-1}\mathcal{C}_n^{(4)}$ |
| $\mathcal{N}_n^{(2)} = \mathcal{C}_n^{(1)} - (n+5-4\nu)\rho^2\mathcal{C}_n^{(2)} + \left(\frac{\rho_\mathrm{i}}{\rho}\right)^{2n+1}\mathcal{C}_n^{(3)} - (n-4+4\nu)\left(\frac{\rho_\mathrm{i}}{\rho}\right)^{2n-1}\mathcal{C}_n^{(4)}$ |

$$\mathcal{N}_n^{(3)} = n(n-1)\mathcal{C}_n^{(1)} - (n+1)[(n+1)(n-2)-2\nu]\rho^2\mathcal{C}_n^{(2)} + (n+1)(n+2)\left(\frac{\rho_\mathrm{i}}{\rho}\right)^{2n+1}\mathcal{C}_n^{(3)} - n[n(n+3)-2\nu]\left(\frac{\rho_\mathrm{i}}{\rho}\right)^{2n-1}\mathcal{C}_n^{(4)}$$
$$\mathcal{N}_n^{(4)} = (n-1)\mathcal{C}_n^{(1)} - (n^2+2n-1+2\nu)\rho^2\mathcal{C}_n^{(2)} - (n+2)\left(\frac{\rho_\mathrm{i}}{\rho}\right)^{2n+1}\mathcal{C}_n^{(3)} + (n^2-2+2\nu)\left(\frac{\rho_\mathrm{i}}{\rho}\right)^{2n-1}\mathcal{C}_n^{(4)}$$
$$\mathcal{N}_n^{(5)} = n^2\mathcal{C}_n^{(1)} - (n+1)(n^2+4n+2+2\nu)\rho^2\mathcal{C}_n^{(2)} + (n+1)^2\left(\frac{\rho_\mathrm{i}}{\rho}\right)^{2n+1}\mathcal{C}_n^{(3)} - n(n^2-2n-1+2\nu)\left(\frac{\rho_\mathrm{i}}{\rho}\right)^{2n-1}\mathcal{C}_n^{(4)}$$
$$\mathcal{N}_n^{(6)} = n\mathcal{C}_n^{(1)} - (n+1)[n-2-2(2n+1)\nu]\rho^2\mathcal{C}_n^{(2)} - (n+1)\left(\frac{\rho_\mathrm{i}}{\rho}\right)^{2n+1}\mathcal{C}_n^{(3)} + n[n+3-2(2n+1)\nu]\left(\frac{\rho_\mathrm{i}}{\rho}\right)^{2n-1}\mathcal{C}_n^{(4)}$$

In the practical computation, we should set a truncation $N_{\mathrm{max}}$ for summing up the terms in Eqs. (13). Figure 2 shows the converge of $\tilde{u}_\rho^{(\mathrm{st})}(\rho,0)$ against $\rho$ for various $N_{\mathrm{max}}$. The convergence becomes worse for $\rho \lesssim 1$ as expected in Eqs. (13). However, it is sufficient to adopt $N_{\mathrm{max}} = 10^3$ even in the $\rho \to 1$ limit. Therefore, we use $N_{\mathrm{max}} = 10^3$ in the following. It should be also noted that the expressions reduce to those for solid spheres [5, 12] when we consider $\rho_\mathrm{i} \to 0$.

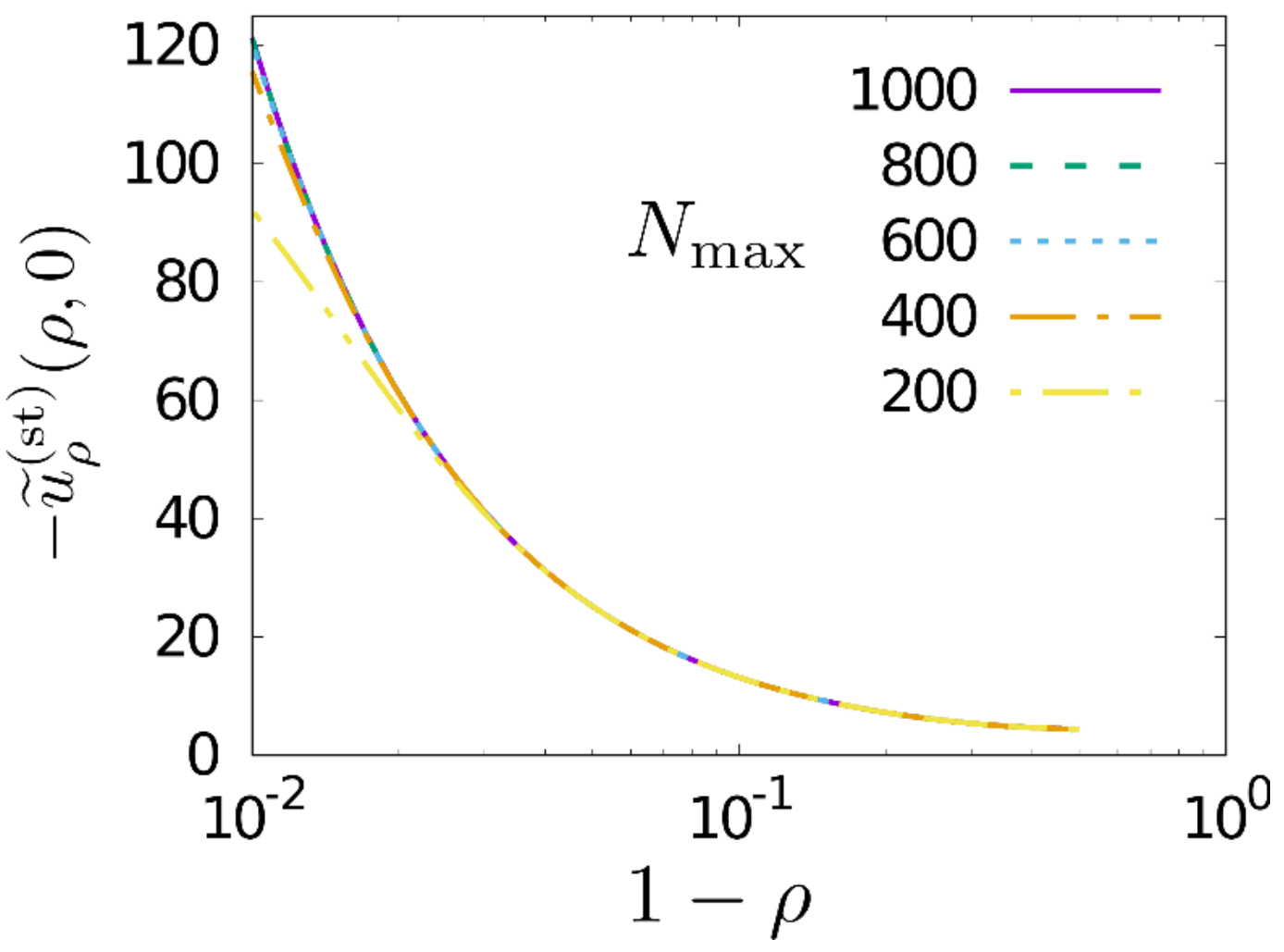

Fig. 2. Convergence of $\tilde{u}_\rho^{(\mathrm{st})}(\rho,0)$ against $\rho$ for various upper limit $N_{\mathrm{max}} = 200, 400, 600, 800$, and 1000 with $\nu = 0.3$ and $\rho_\mathrm{i} = 0.5$.

**4. Results.** Let us show the results obtained from the theory. To visualize the results, the (dimensionless) magnitude of the displacement $\tilde{u}^{(\mathrm{st})}$ and the (dimensionless) second principal stress difference (principal stress difference in $(\rho,\theta)$ plane) $\Delta\sigma^{(\mathrm{st})}$ are often used [17]:

$$\tilde{u}^{(\mathrm{st})} \equiv \sqrt{\left(\tilde{u}_\rho^{(\mathrm{st})}\right)^2 + \left(\tilde{u}_\theta^{(\mathrm{st})}\right)^2}, \qquad \Delta\sigma^{(\mathrm{st})} \equiv \sqrt{\left(\tilde{\sigma}_{\rho\rho}^{(\mathrm{st})} - \tilde{\sigma}_{\theta\theta}^{(\mathrm{st})}\right)^2 + 4\left(\tilde{\sigma}_{\rho\theta}^{(\mathrm{st})}\right)^2}. \tag{14}$$

Figure 3 shows the magnitude plots of $\tilde{u}^{(\mathrm{st})}$ and $\Delta\sigma^{(\mathrm{st})}$ for $\nu = 0.3$ and $\rho_\mathrm{i} = 0.5$. Both quantities are enhanced in the vicinity of the loading points.

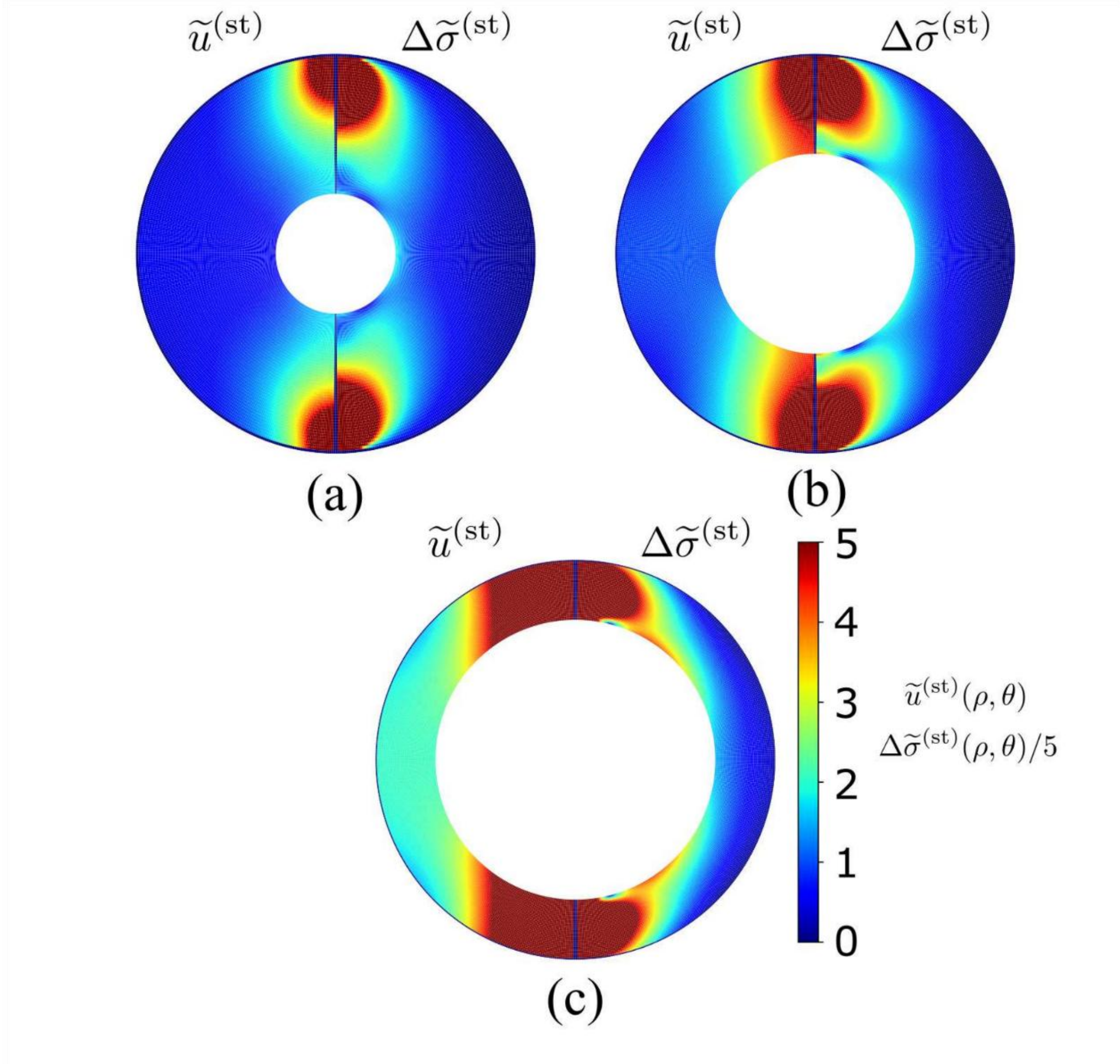

Fig. 3. The magnitude of (left) the deformation $\tilde{u}^{(\mathrm{st})}$ and (right) the second principal stress difference $\Delta\sigma^{(\mathrm{st})}$ for (a) $\rho_{\mathrm{i}} = 0.3$, (b) 0.5, and (c) 0.7 when we fix $\nu = 0.3$.

We can also check the behaviors of the compression and extension stresses on the loading line $\theta = 0$. Figure 4 shows how both stresses deviate from those for a solid sphere ($\rho_{\mathrm{i}} = 0$). First, the divergent behaviors of both stresses are observed in the vicinity of the outer surface ($\rho \to 1$), which is common with those for a solid sphere. Second, the compression stress decreases to zero and the extension stress becomes larger in the vicinity of the inner surface ($\rho \gtrsim \rho_{\mathrm{i}}$). Figure 4 clearly shows how those stresses deviate from those of a solid sphere [12]. The compression stress of a hollow sphere is the same in most regions ($\rho_{\mathrm{i}} < \rho < 1$) compared to those of a solid sphere, and decrease sharply as the inner diameter is approached ($\rho \to \rho_{\mathrm{i}}$). On the other hand, the extension stress is smaller than that of a solid sphere in the region between the outer and inner diameters. The stress is then higher as the inner diameter is approached.

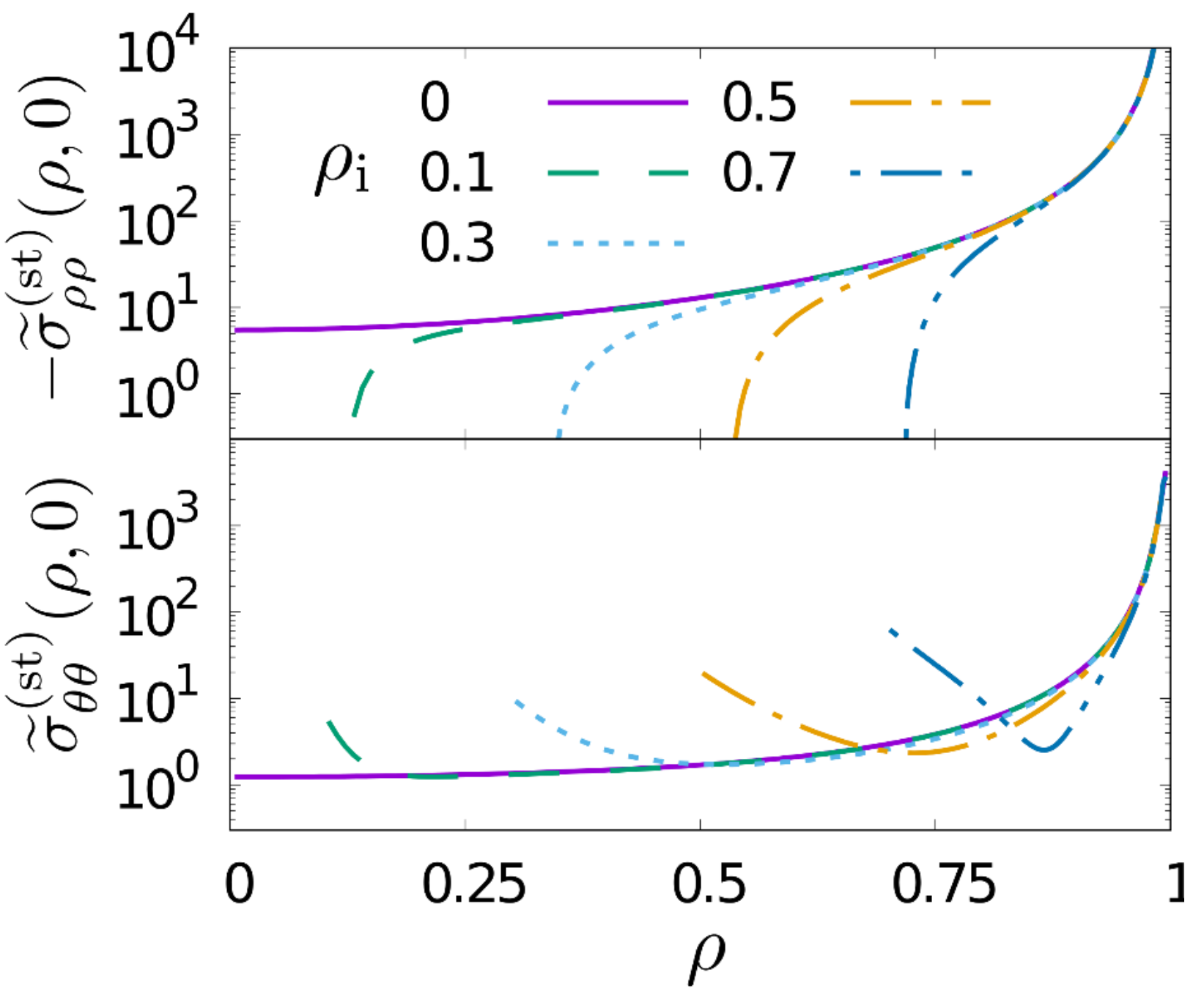

Fig. 4. Plots of (upper) the compression stress $-\tilde{\sigma}^{(\mathrm{st})}_{\rho\rho}$ and (bottom) the extension stress $\tilde{\sigma}^{(\mathrm{st})}_{\theta\theta}$ on the loading line ($\theta = 0$) for various $\rho_{\mathrm{i}}$ with $\nu = 0.3$. The solid lines represent those for a solid sphere.

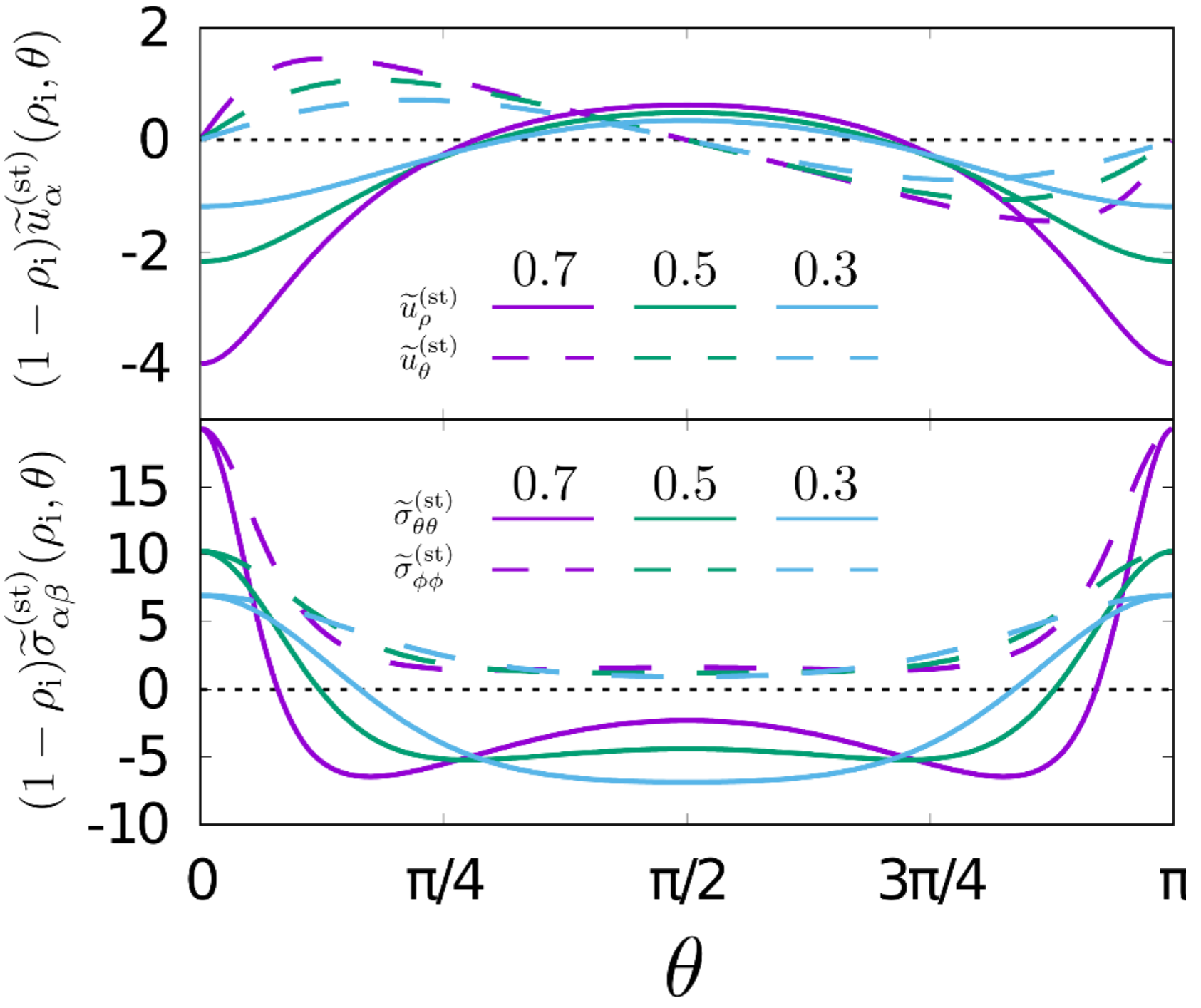

Fig. 5. Plots of (upper) the deformation and (bottom) the stress components against $\theta$ at the inner surface for $\rho_{\mathrm{i}} = 0.3$, 0.5, and 0.7.

Let us observe the behaviors of $\tilde{u}^{(\mathrm{st})}_{\alpha}$ ($\alpha = \rho, \theta$) and $\tilde{\sigma}^{(\mathrm{st})}_{\alpha\beta}$ ($(\alpha,\beta) = (\theta,\theta),(\phi,\phi)$) at the inner edge ($\rho = \rho_{\mathrm{i}}$). Figure 5 shows their behaviors for $\rho_{\mathrm{i}} = 0.3$, 0.5, and 0.7. Here, we do not show $\tilde{\sigma}^{(\mathrm{st})}_{\rho\rho}(\rho_{\mathrm{i}},\theta)$ and $\tilde{\sigma}^{(\mathrm{st})}_{\rho\theta}(\rho_{\mathrm{i}},\theta)$ because they are zero by definition of the boundary condition (1). Interestingly, the values at $\theta = \pi/2$ can be approximately scaled as $\tilde{u}_\alpha^{(\mathrm{st})}, \tilde{\sigma}_{\alpha\beta}^{(\mathrm{st})} \propto (1-\rho_\mathrm{i})^{-1}$. This can be understood when we consider the thin shell limit of the system. Let us consider the situation where the hole radius $R_\mathrm{i}$ is sufficiently large, that is, $R_\mathrm{o} - R_\mathrm{i} \ll 1$. In this case, let us define the shell width

$$\delta \equiv 1 - \rho_\mathrm{i} \ (\ll 1). \tag{15}$$

For example, the radial displacement at the surface in $\delta \to 0$ limit become

$$\tilde{u}_\rho^{(\mathrm{st})} = \tilde{u}_\rho^{(\mathrm{st},-1)}(\theta)\frac{1}{\delta} + \tilde{u}_\rho^{(\mathrm{st},0)}(\theta) + \mathcal{O}(\delta), \tag{16a}$$

$$\tilde{u}_\rho^{(\mathrm{st},-1)} \equiv -\sum_{n=0,2,4,\cdots}^{\infty} \frac{(2n+1)(n^2+n-1+\nu)}{2(n-1)(n+2)(1+\nu)} P_n(\cos\theta), \tag{16b}$$

$$\tilde{u}_\rho^{(\mathrm{st},0)}(\theta) \equiv \sum_{n=0,2,4,\cdots}^{\infty} \frac{(2n+1)\nu}{2(1+\nu)} P_n(\cos\theta), \tag{16c}$$

Note that the similar discussions are available on $\tilde{u}_\theta^{(\mathrm{st})}$ and $\tilde{\sigma}_{\alpha\beta}^{(\mathrm{st})}$. This indicates that the above scaling is satisfied, at least, for $\rho_\mathrm{i} \lesssim 1$.

In addition, $\tilde{\sigma}_{\theta\theta}^{(\mathrm{st})}(\rho_\mathrm{i}, \theta)$ has a peak when $\rho_\mathrm{i}$ is not small as shown in Fig. 5. The second term in Eq. (13e) is a decreasing function of $\theta$ in the range of $0 \le \theta \le \pi/2$ and becomes zero at $\theta = \pi/2$. The second term survives even near $\theta = \pi/2$ and its value becomes larger as $\rho_\mathrm{i}$ becomes large. This is the origin of the existence of peaks in the middle of $0 < \theta < \pi/2$.

**Conclusion and Discussion.** In this study, we revisited the stress analysis of a three-dimensional elastic hollow sphere under uniaxial compression. Using linear elastodynamic theory, we introduced scalar and vector potentials to simplify the problem and determined them under the boundary condition. The deformation and stress fields were derived as inverse Laplace forms, with static solutions explicitly expressed as infinite series involving Legendre polynomials. A detailed discussion of the convergence of these series confirmed their validity. Graphical representation of the deformation and principal stress difference provided insights into the stress distribution. Notably, we observed a peak in the circumferential stress $\sigma_{\theta\theta}$ at the inner surface when the inner radius became sufficiently large, highlighting the significant influence of geometric factors. These findings contribute to a deeper understanding of stress behavior in hollow spherical structures and have implications for materials and structural design.

Our methodology can be extended to dynamic problems as well. Indeed, by evaluating the inverse Laplace transform expressions, it is possible to determine displacements and stresses at any given time and location [8, 9, 11, 12]. In such cases, phenomena such as longitudinal waves, transverse waves, and Rayleigh waves traveling along the surface are known to appear [2]. However, as mentioned in the main text, it is crucial to skillfully evaluate the residues and carefully consider which waves have arrived based on the time and location. This makes it likely that the results will need to be categorized into multiple cases, significantly increasing the complexity of the analysis.

**Acknowledgment.** The authors thank Hisao Hayakawa for his valuable comments and suggestions. This work is supported by the Grant-in-Aid of MEXT for Scientific Research (Grant No. JP24K06974, No. JP24K07193, and No. JP25K01063).

**Conflict of Interests.** All authors declare that they have no conflicts of interest.

**Author Contributions.** All authors contributed to the study conception and design. The main derivation and analysis were performed by Satoshi Takada. The figures were prepared by Shintaro Hokada. The first draft of the manuscript was written by Satoshi Takada and all authors commented on previous versions of the manuscript. All authors read and approved the final manuscript.